\documentstyle[sprocl]{article}
\input{epsf}
\input{psfig}
\def\be{\begin{equation}}
\def\ee{\end{equation}}
\def\bea{\begin{eqnarray}}
\def\eea{\end{eqnarray}}

\begin{document}

\title{HYPERON PRODUCTION ASYMMETRIES IN 500 GeV/$c$ PION 
NUCLEUS INTERACTIONS}

\author{J. SOLANO}
\author{representing the}
\address{}
\author{FERMILAB E791 COLLABORATION}
\author{and J. MAGNIN, F.R.A. SIM\~AO}
\address{Centro Brasileiro de Pesquisas F\'{\i}sicas - CBPF 
Rio de Janeiro, Brazil.}
\maketitle

\abstracts{We present a preliminary study from Fermilab experiment E791 of $\Lambda^0 /\overline{\Lambda}^0$, $\Xi^- /\Xi^+ $ 
and $\Omega^- /\Omega^+ $ production asymmetries from $\pi^-$ nucleus
interactions at 500 Gev/$c$ .The production asymmetries 
for these particles are studied as a function of $x_F$ 
and $p_{T}^2$. We observed an asymmetry in the target fragmentation region 
for $\Lambda^0$'s larger than that for $\Xi$'s, suggesting diquark effects. 
The asymmetry for $\Omega$'s is significatively smaller than for the other 
two hyperons consistent with the fact that $\Omega$'s do not share valence quarks with either the pion or the target particle. In the beam fragmentation
region, the asymmetry tends to 0.1 for both $\Lambda^0$'s and $\Xi$'s. The asymmetries vs $p_T^2$ are approximately constant for the three strange 
baryons under study.}

Leading particle effects in  charm hadron production, which are manifest 
as an enhancement in the production rate of particles which share 
valence quarks over that of particles with no valence quarks in common with
the initial hadrons, have been extensively studied in the recent years from 
both the experimental and theoretical point of view.
The same type of leading effects are expected to occur in light hadron 
production. Indeed, there is some evidence of asymmetries in 
$\Lambda^0 /\overline{\Lambda}^0$ production in $\pi^-$Cu interactions 
at 230 GeV/$c$ \cite{accmor1} and in $\Lambda^0 /\overline{\Lambda}^0$ and 
$\Xi^-/\Xi^+$ production in 250 GeV/$c$ $\pi^-p$ interactions \cite{bogert}. 
Some additional evidence for $\Lambda^0 /\overline{\Lambda}^0$ asymmetry 
can be found in Ref. 3, but in general light hadron 
production asymmetries in $\pi^-p$ interactions are not systematically 
studied. \\ 

As a byproduct of our charm program in Fermilab Experiment E791 we
collected a large sample of $\Lambda^0 /\overline{\Lambda}^0$, 
$\Xi^- /\Xi^+ $ and $\Omega^- /\Omega^+ $ which was used to measure 
the particle/anti-particle production asymmetries reported in this paper. 
The $x_F$ range covered by our experiment is 
$-0.12 \leq x_F \leq 0.12$ allowing for the study of baryon/anti-baryon 
production asymmetries in the target ($x_F<0$) and beam ($x_F>0$) 
fragmentation regions. {\em A priori}, strong differences are expected for the 
asymmetries observed in each region where there is a different content of 
valence quarks in the produced particles relative to the anti-particles with 
respect to the target and beam hadrons. Thus a growing asymmetry with 
$\left| x_F \right|$ is expected in $\Lambda^0/\overline{\Lambda}^0$ 
production in the target fragmentation region. A smaller or no 
asymmetry is expected in the beam fragmentation region, in which both 
$\Lambda^0$ and $\overline{\Lambda}^0$ share one valence quark 
with the pion. For $\Xi^- /\Xi^+ $ production a growing asymmetry 
with $\left| x_F \right|$ is expected in both regions since the 
$\Xi^-$ shares one valence quark with the $\pi^-$ as well as with the 
target particles ($p$ and $n$) whereas the $\Xi^+$ share none. A zero 
asymmetry is expected for $\Omega^- /\Omega^+$ which have no valence 
quarks in common with either the target or the beam.\\

In the E791 experiment, data were recorded from 500 GeV/$c$ $\pi^-$ 
interactions in five thin foils (one platinum and four diamond) 
separated by gaps of 1.34 to 1.39 cm. The E791 spectrometer \cite{vinte} is 
an upgrade  of apparatus used in Fermilab charm experiments E516, E691 and 
E769. It is a large-acceptance, two-magnet spectrometer with
eight planes of multiwire proportional chambers (MWPC) and six
planes of silicon microstrip detectors (SMD) for beam tracking. 
Downstream of the target are a 17-plane
SMD system for track and vertex reconstruction,
35 drift chamber planes, two MWPC's, two multicell threshold
Cherenkov counters, electromagnetic and hadronic calorimeters and a
muon detector. An important element of the experiment was its extremely 
fast data acquisition system \cite{viteum} which was combined with a 
very open trigger to record a data sample of 20 x 10$^9$ interactions.\\

$\Lambda^0$'s were studied via their $p\pi^-$ decay mode.  The distance 
of closest approach between the proton and $\pi^-$ tracks must be less that 
0.7 cm to be considered as candidates for $\Lambda^0$ decay and the candidate 
$\Lambda^0$ should have an invariant mass between 1.101 and 1.127 GeV/$c^2$. 
In addition, the ratio of the momentum of the proton to the pion is required 
to be larger than 2.5. The reconstructed $\Lambda^0$ decay vertex formed by 
the two tracks is required to be downstream of the last target. In order to avoid $\Lambda^0$'s coming from $\Xi$ decay, candidates must have an impact parameter with respect to the primary vertex less than 0.3 cm for $\Lambda^0$'s decaying within the first 20 cm and 0.4 cm for $\Lambda^0$'s decaying 
downstream of 20 cm from the target.  The reconstructed mass distribution of 
the obtained sample was fitted using a binned maximum likelihood with a 
Gaussian plus a linear background. We obtained 2,571,662 $\pm$ 3,057 $\Lambda^0$'s and 1,668,950 $\pm$ 2,627 $\overline{\Lambda}^0$'s, from approximately 6$\%$ of the total data sample recorded in the experiment.\\

$\Xi$'s are selected via the decay mode $\Xi \rightarrow \Lambda^0 \pi$ 
and at the same time $\Omega$'s are selected in the channel $\Omega 
\rightarrow \Lambda^0 K$. Starting with candidate $\Lambda^0$'s 
we add a third track as a possible pion/kaon daughter. 
The cuts applied to the $\Lambda^0$ are the same quoted above except 
for the impact parameter cut, which is not applied in this case, and 
the third track must meet the following requirements: the track cannot be 
one of those that made the $\Lambda^0$, the invariant mass of the candidate 
$\Xi$ (calculated from the mass of the $\Lambda^0$ and the net momentum of 
the two decays tracks plus the third track) must be between 1.290 and 1.350 GeV/$c^2$ for $\Xi$'s and 1.642 and 1.702 GeV/$c^2$ for the $\Omega$'s. 
In addition, $\Xi$'s and $\Omega$'s decaying upstream of the SMD and
downstream of the $\Lambda^0$ are not allowed and the efficiency of SMD's
for the candidate hyperon track must be larger than 0.75. An additional cut 
is applied for the third track: the Cherenkov probability of this track to be a kaon mut be larger than 0.13 if its momentum is smaller than 40 GeV/c and 
larger than 0.018 if its momentum is larger than 40 GeV/c. As with the 
lambda sample, we fitted the cascade and omega samples using maximum likelihood 
with a Gaussian plus a linear background. From approximately $2/3$ of the 
total E791 data sample we found 570,535 $\pm$ 1,386 $\Xi^-$; 401,156 $\pm$ 
1,166 $\Xi^+$; 7,401 $\pm$ 132 $\Omega^-$ and 6,598 $\pm$ 97 $\Omega^+$ 
after background substraction.\\

\begin{figure}[ht]

\begin{center}

\begin{minipage}{2.5in}
\psfig{figure=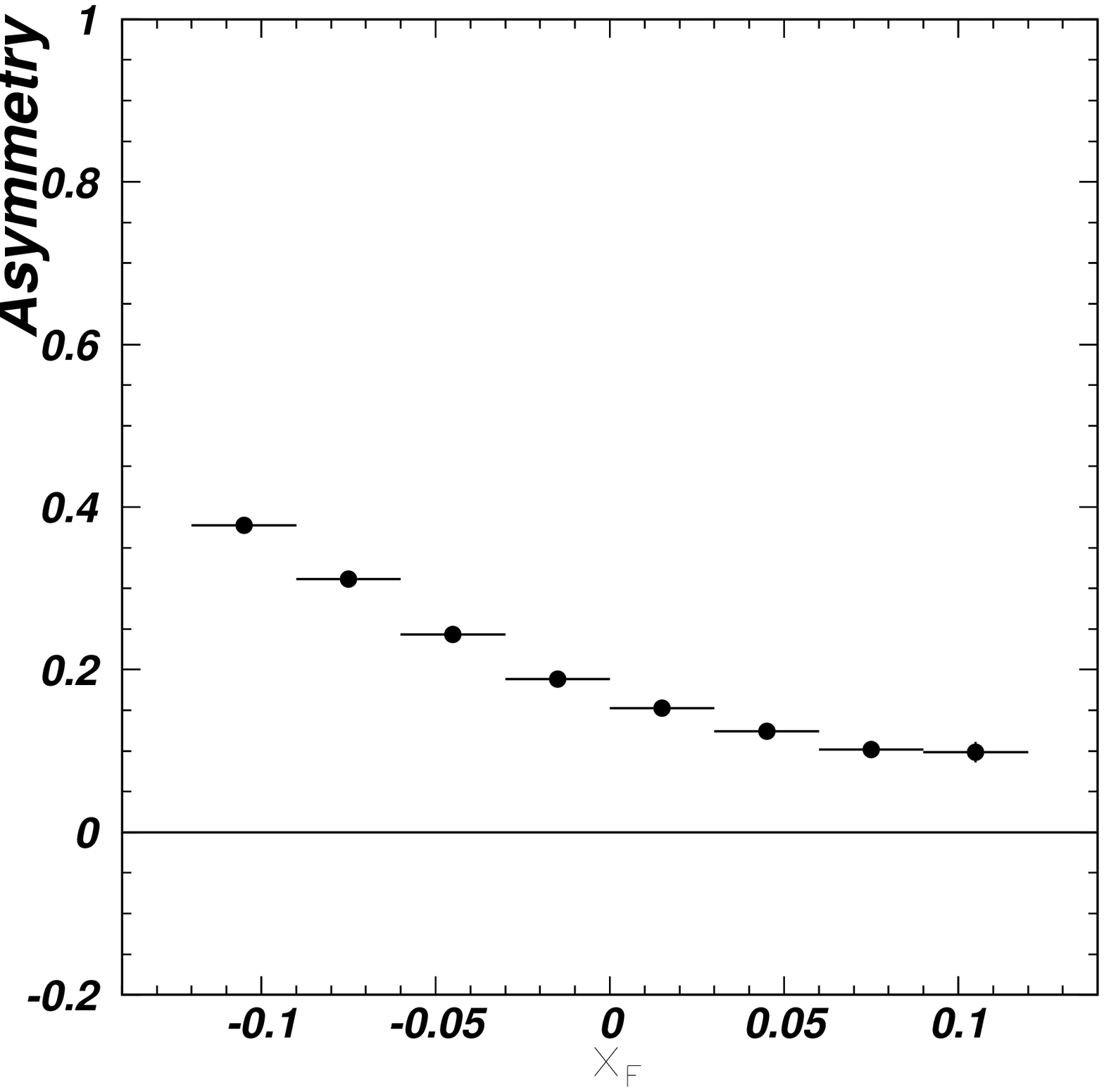,height=2.0in}
\end{minipage}
\begin{minipage}{7.5in}
\psfig{figure=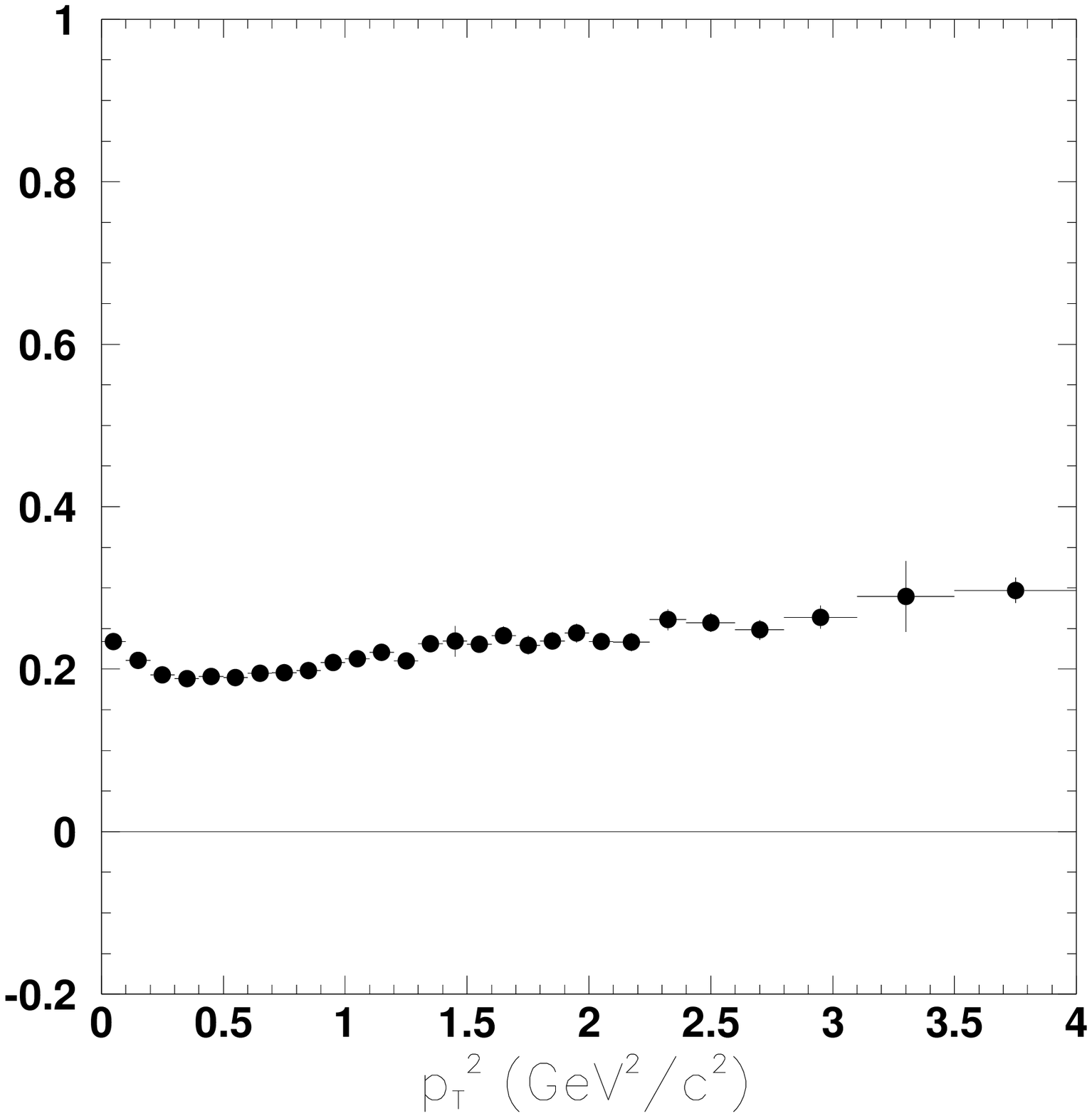,height=2.0in}
\end{minipage}
\begin{minipage}{2.5in}
\psfig{figure=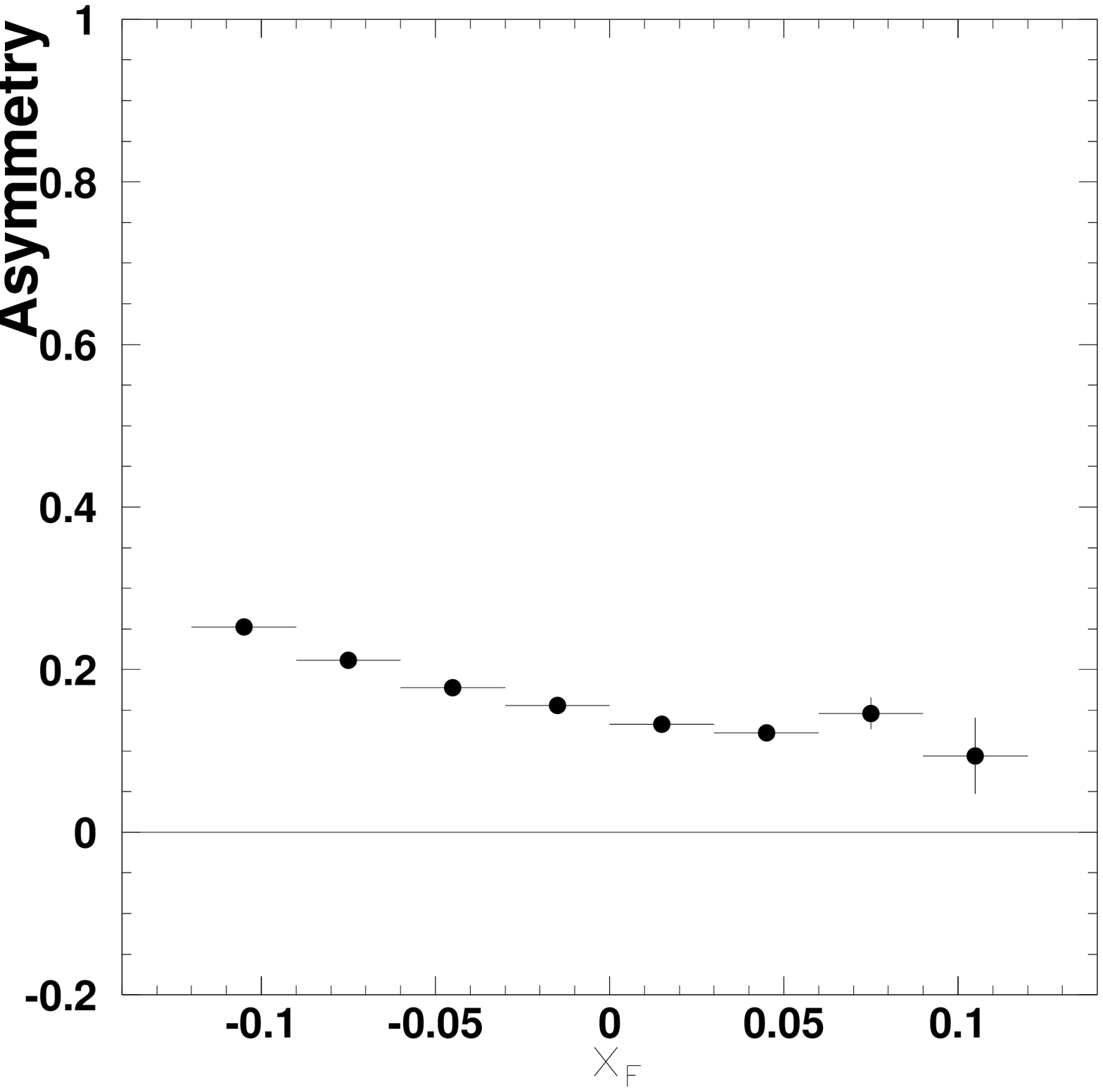,height=2.0in}
\end{minipage}
\begin{minipage}{7.5in}
\psfig{figure=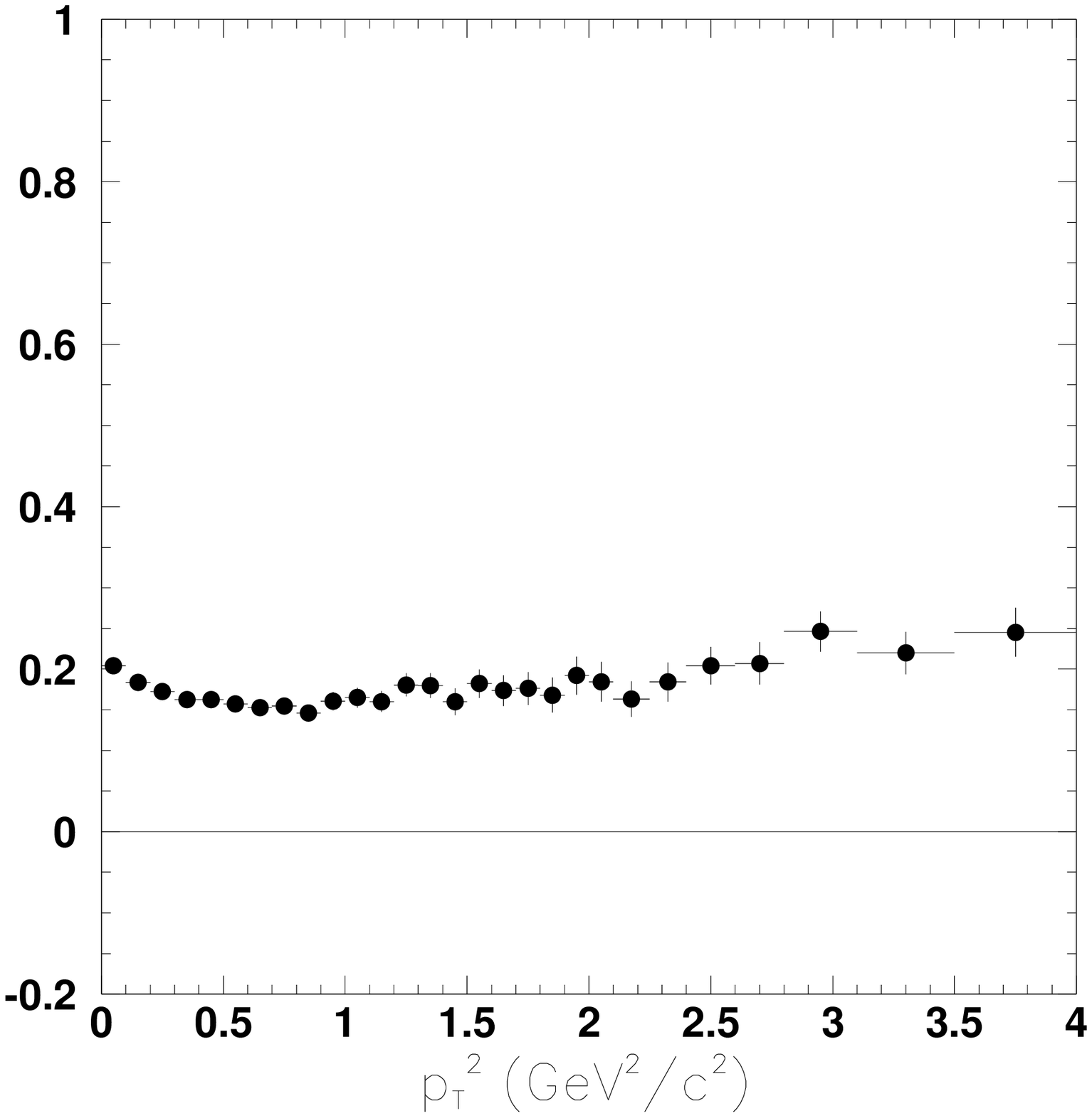,height=2.0in}
\end{minipage}
\begin{minipage}{2.5in}
\psfig{figure=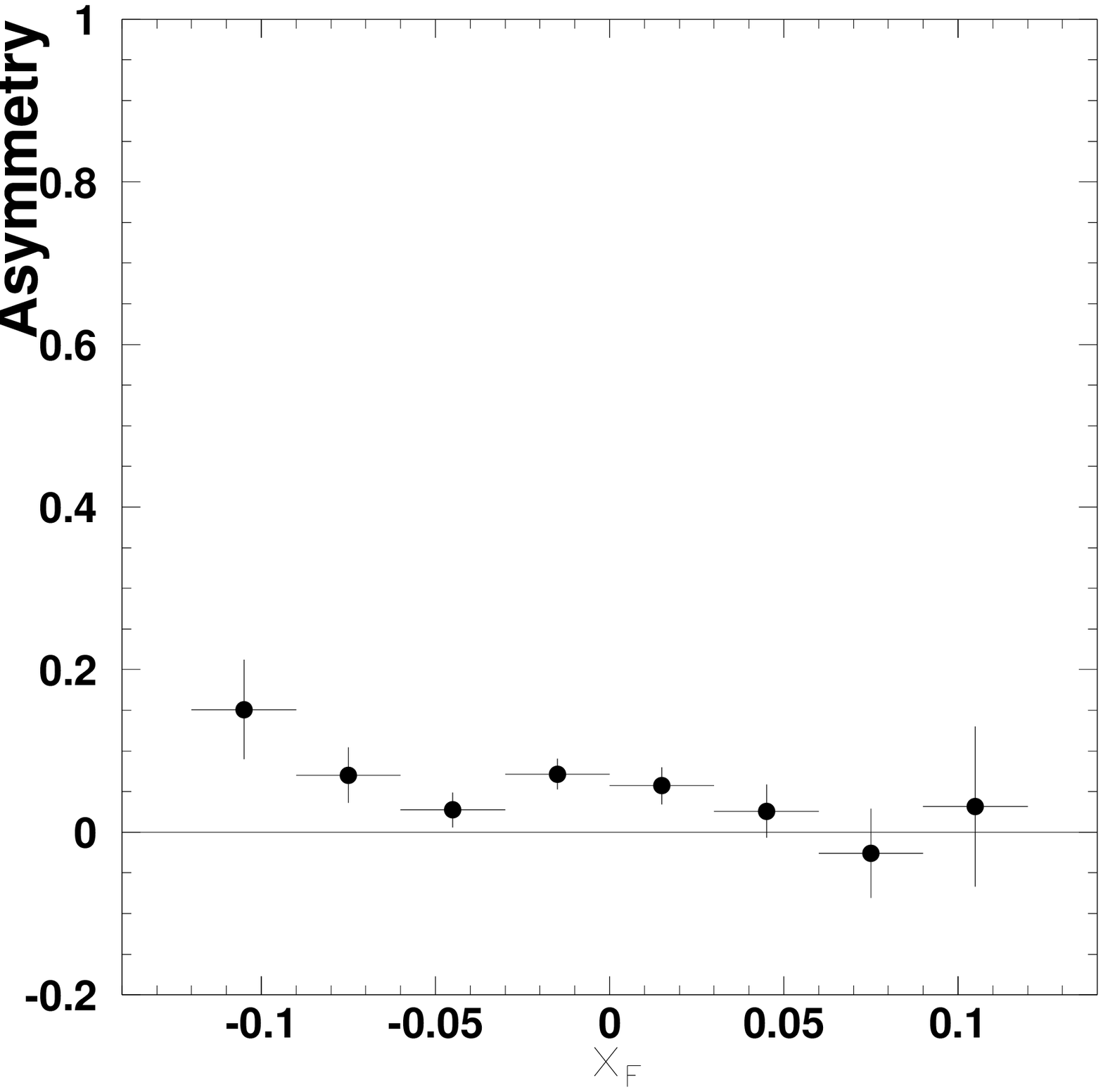,height=2.0in}
\end{minipage}
\begin{minipage}{2.5in}
\psfig{figure=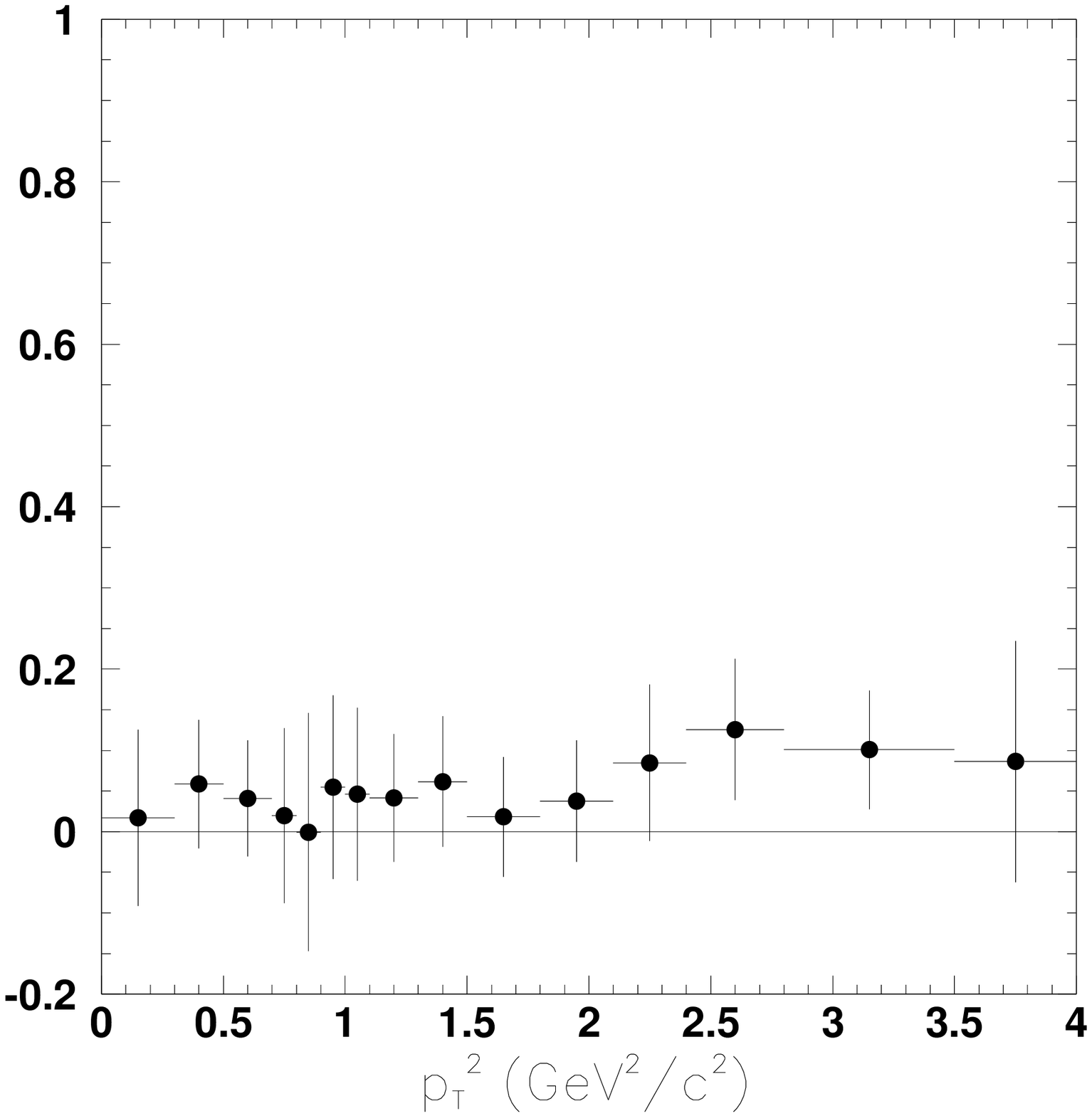,height=2.0in}
\end{minipage}

\end{center}

\caption{Preliminary results for 6$\%$ of the E791 data for $\Lambda$ and 
2/3 for $\Xi$ and $\Omega$. Statistical errors only. 
$\Lambda^0/\overline{\Lambda}^0$ (top), $\Xi^-/\Xi^+$ (middle) and 
$\Omega^-/\Omega^+$ (bottom) asymmetries {\it vs.} $x_F$ (left) 
and $p_{T}^2$ (right).}

\label{total}

\end{figure}

For each $x_F$ and $p_T^2$ bin we define an asymmetry parameter A as
\begin{equation}
A\left(B/\overline{B}\right) \equiv \frac{N_{B} - N_{\overline {B}}}
{N_{B} + N_{\overline {B}}} \; ,
\end{equation}
where $N_B$ ($N_{\overline {B}}$) is the number of baryons (anti-baryons)
in the bin. Monte Carlo studies done with PYTHIA \cite {lund} for 
$\Lambda^0$ and $\overline{\Lambda}^0$ have shown that efficiencies 
for both particle and anti-particle are the same therefore no corrections are 
nedeed in this case. For $\Xi$'s and $\Omega$'s we are finishing 
the process of MC generation in order to study efficiency. The obtained asymmetries are shown as functions of $x_F$ and $p_T^2$ 
in Fig. \ref{total} for $\Lambda^0/\overline{\Lambda}^0$, $\Xi^-/\Xi^+$ and $\Omega^-/\Omega^+$. The total asymmetries measured in the interval $-0.12 \leq x_F \leq 0.12$ 
and in the backward and forward $x_F$ regions are shown in Table 1.\\ 
In the backward $x_F$ region we observe an asymmetry growing with 
$\left| x_F \right|$ for both $\Lambda^0$'s and $\Xi$'s and an 
approximately constant and small asymetry for $\Omega$'s. 
As can be seen in the figures, the asymmetry in the backward region 
is higher as more valence quarks are shared between the produced 
particle and protons and neutrons in the target. 
The higher asymmetry observed for the $\Lambda^0$'s may be due 
to the $ud$ diquark shared by the produced $\Lambda^0$ and the 
protons and neutrons in the target. In the forward region the asymmetry for 
both $\Lambda^0$'s and $\Xi$'s is approximately constant. For the 
$\Xi^-/\Xi^+$ the observed asymmetry is compatible with the fact that 
the $\Xi^-$ shares one valence quark with the incident $\pi^-$ whereas 
the $\Xi^+$ has no valence quarks in common with the $\pi^-$, so some 
leading effect must be present. 

\begin{table}
\caption{Hyperon asymmetries. Only statistical errors are shown.}
\begin{center}
\begin{tabular}{|l|l|l|l|} \hline\hline
 & $-0.12\leq x_F \leq 0.12$ & $-0.12 < x_F < 0$ & 
$0 \leq x_F < 0.12$\\ \hline

$A(\Lambda^0/\overline{\Lambda}^0)$ & $0.213 \pm 0.001$ & 
$0.246 \pm 0.001$ & $0.138 \pm 0.002$ \\

$A(\Xi^-/\Xi^+)$ & $0.174 \pm 0.002$ & 
$0.186 \pm 0.002$ & $0.131 \pm 0.004$ \\

$A(\Omega^-/\Omega^+)$ & $0.057 \pm 0.012$ & 
$0.058 \pm 0.015$ & $0.056 \pm 0.017$ \\ \hline \hline 
\end{tabular}
\end{center}
\end{table}


The asymmetry measured for the $\Lambda^0/\overline{\Lambda}^0$ 
in the forward region can not be explained by a leading-particle effect 
because both $\Lambda^0$ and $\overline{\Lambda}^0$ have one valence 
quark in common with the pion.\\
For the three strange baryons we observe an approximately flat 
asymmetry, of the order of 0.2 for $\Lambda$'s and $\Xi$'s and very 
small for $\Omega$'s, in the complete $p_T^2$ interval studied. \\

\section*{Acknowledgments}
J.M. is supported by FAPERJ, Funda\c{c}\~ao de Amparo \`a Pesquisa 
do Estado de Rio de Janeiro.
\section*{References}

\end{document}